\documentclass[nofootinbib,superscriptaddress,twocolumn,showpacs,preprintnumbers,pre,aps]{revtex4-1}

\usepackage{graphicx}
\usepackage{bm}
\usepackage{amsmath}
\usepackage{amssymb}
\usepackage{color}
 
\DeclareMathOperator{\Det}{Det}

\begin{document}

\title{Relaxation dynamics of a compressible bilayer vesicle containing highly viscous fluid}

\author{T. V. Sachin Krishnan}

\affiliation{
Department of Chemistry, Graduate School of Science and Engineering,
Tokyo Metropolitan University, Tokyo 192-0397, Japan}

\affiliation{
Department of Physics, Indian Institute of Technology Madras,
Chennai, 600036, India}

\author{Ryuichi Okamoto}

\affiliation{
Department of Chemistry, Graduate School of Science and Engineering,
Tokyo Metropolitan University, Tokyo 192-0397, Japan}

\author{Shigeyuki Komura}\email{komura@tmu.ac.jp}
\affiliation{
Department of Chemistry, Graduate School of Science and Engineering,
Tokyo Metropolitan University, Tokyo 192-0397, Japan}

\date{\today}

\begin{abstract}
We study the relaxation dynamics of a compressible bilayer vesicle with an asymmetry in the 
viscosity of the inner and outer fluid medium.
First we explore the stability of the vesicle free energy which includes a coupling between the 
membrane curvature and the local density difference between the two monolayers.  
Two types of instabilities are identified: a small wavelength instability and a larger wavelength instability.
Considering the bulk fluid viscosity and the inter-monolayer friction as the dissipation sources, 
we next employ Onsager's variational principle to derive the coupled equations both
for the membrane and the bulk fluid. 
The three relaxation modes are coupled to each other due to the bilayer and the spherical 
structure of the vesicle.
Most importantly, a higher fluid viscosity inside the vesicle shifts the cross-over mode between the 
bending and the slipping to a larger value.  
As the vesicle parameters approach toward the unstable regions, the relaxation dynamics is 
dramatically slowed down, and the corresponding mode structure changes significantly.
In some limiting cases, our general result reduces to the previously obtained relaxation rates. 
\end{abstract}

\maketitle

\section{Introduction}
\label{sec:introduction}

Macromolecules such as proteins, nucleic acids, and polysaccharides are present in high concentration in cells and physically occupy up to 40\% of the total cell volume~\cite{Alberts08, Phillips12}. 
This feature, referred to as ``macromolecular crowding", has a significant effect on important processes taking place inside the cell.  
This is due to the fact that any reaction which depends on the available volume and the diffusivity 
of reactants is affected by crowding~\cite{Ellis03}.
The increase in the effective viscosity of the crowded environment is considered to be 
responsible for the observed anomalous diffusion of proteins inside cells~\cite{Zimmerman93,Banks05}.
It is known that crowding influences the conformation, stability, and the function of 
nucleic acids~\cite{Miyoshi08} and proteins~\cite{Kuznetsova14}. 
However, most of the \textit{in-vitro} biochemical and biophysical studies involving macromolecules are carried out in highly dilute environments which do not reflect the natural conditions inside the cells~\cite{Ellis01}.

Artificial cells with crowded internal environment are being developed in order to study the macromolecules in their physiological conditions~\cite{Kurihara11, Kuruma15}.
Giant unilamellar vesicles (GUVs), which are cell-sized spherical shells consisting of a lipid bilayer, are commonly 
used as containers for artificial cells in which many synthetic chemical reactions take place. 
In a recent paper by Fujiwara and Yanagisawa~\cite{Fujiwara14}, they used GUVs filled with macromolecules
in order to investigate the effect of higher viscosity inside the vesicles.
When these GUVs were subjected to osmotic pressure, they exhibited different shape deformations depending
on the inside viscosity. 
This result suggests that the asymmetry in the viscosity affects the shape dynamics of the bounding lipid 
membrane~\cite{Fujiwara14}.

One of the early quantitative works on dynamics of membranes is by Brochard and Lennon who studied the flickering phenomena in red blood cells (RBCs)~\cite{Brochard75}.
They identified the observed flickering of RBC as equilibrium thermal fluctuations of flexible lipid bilayers~\cite{Boss12}. 
Analysis of flickering phenomena is a useful method to estimate elastic properties of fluid membranes.
Furthermore, measurements of membrane flickering can be used to identify diseases affecting biological cells. 
On the other hand, recent experiments revealed that the flickering is also controlled by active 
ATP-dependent processes taking place inside the cell~\cite{Betz09, Park10, Turlier16}.

Even though the relaxation dynamics of fluid vesicles has been studied theoretically for several 
decades~\cite{Schneider84,Milner87,Miao02,Mell15,Arroyo09}, the effect of fluid viscosity difference 
between inside and outside of the vesicle has not received much attention.
Seki and Komura explicitly took into account the viscosity difference, and obtained an expression for 
the bending relaxation mode of quasi-spherical vesicles~\cite{Komura93,Seki95,Komura96}. 
However, since the bilayer nature of the vesicle was not considered in their calculations, the relaxation 
of the lipid densities on the monolayers mediated by the inter-monolayer friction was not taken into account.
Relaxation mode governed by this inter-monolayer friction (called the ``slipping mode") affects the small 
wavelength dynamics of planar membranes~\cite{Evans92,Seifert93,Okamoto16,Fournier15}.

In this work, we investigate the relaxation dynamics of a vesicle by explicitly taking into account 
the effect of fluid viscosity contrast between inside and outside of the vesicle. 
We consider a quasi-spherical, single-component, unilamellar, and compressible bilayer vesicle. 
The vesicle shape deformation and the difference in the local densities of the monolayers 
are governed by the bending rigidity and the membrane compressibility, respectively.
The dissipations arise from the viscosity of the bulk fluids and the inter-monolayer friction.
We employ Onsager's variational principle to derive the relaxation equations for the system that 
can be described by low Reynolds number hydrodynamics~\cite{Doi11,Doi13}. 
We show that the bending and the slipping modes are coupled to each other due to the bilayer architecture. 
The viscosity contrast results in slowing down of the bending mode, and it significantly affects the 
slowest relaxation mode. 
We also discuss the stability of the vesicle free energy, and identify two distinct instabilities.
One of them is associated with the slowing down of the largest wavelength relaxation mode, while the other is 
related to the slowing down of the shorter modes.

In previous works, exact expressions and approximations for the coupled relaxation modes of a 
compressible vesicle have been discussed~\cite{Miao02,Mell15}.
Our goal is different from these works in that we account for the viscosity asymmetry between 
the inside and the outside fluid media.
In addition, we discuss the conditions for stability of the vesicle free energy.
Our calculation also follows a different method based on Onsager's variational principle
in order to derive the appropriate dynamical equations.
We also comment that the effect of different viscosities on the dynamics of vesicles under 
shear flow was studied previously~\cite{Keller82,Kantsler05, Kantsler06}. 
It was found that such a viscosity contrast causes a transition between tank-treading and tumbling 
motions~\cite{Kantsler05,Kantsler06}. 
While vesicles under flow in these works attain one of the non-equilibrium steady states in the long time limit,
vesicles in our study eventually relax to a mechanical equilibrium described by a perfect sphere.

The outline of this paper is as follows.
Section~\ref{sec:model} introduces the parametrization to specify the configuration of the vesicle and 
its free energy functional.
In Sec.~\ref{sec:statics}, we examine the statics of the system near equilibrium, and derive the conditions for thermodynamic stability.
In Sec.~\ref{sec:dynamics}, we construct the Rayleighian functional and derive the equations for the 
relaxation dynamics by extremalizing the Rayleighian.
Section~\ref{sec:results} presents the main results of the work including the effect of viscosity contrast on the relaxation dynamics of the bilayer vesicle. 
Finally, some discussion on the relation of our work to the previous theoretical calculations are provided in 
Sec.~\ref{sec:discussions}.

\section{Free energy of a compressible bilayer vesicle}
\label{sec:model}

\subsection{Variables}

\begin{figure}[tbh]
\centering
\includegraphics[scale=0.65]{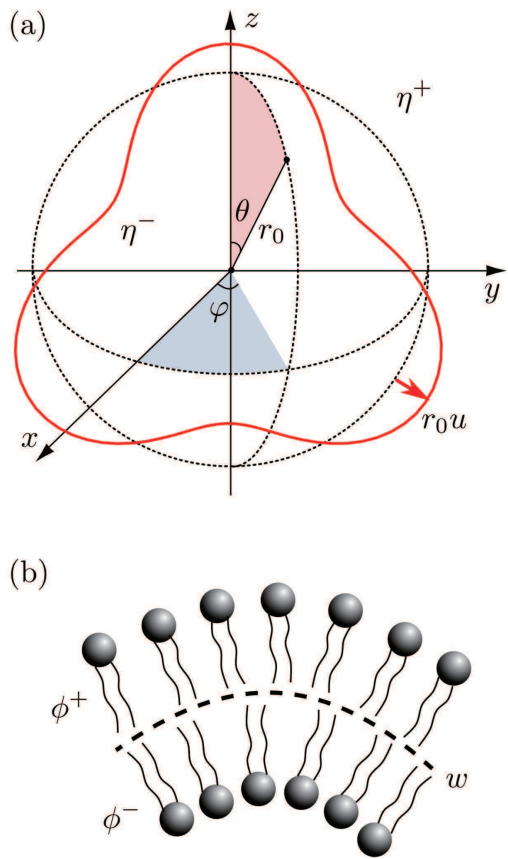}
\caption{(a) Schematic picture of a vesicle showing the reference sphere (dashed line) of radius $r_0$  
with fluids of viscosity $\eta^-$ inside and $\eta^+$ outside. 
Solid red line represents a deformed vesicle configuration whose shape is parametrized with $u(\theta, \varphi)$. 
(b) Cross-section of a curved region in the bilayer showing the lipid molecules in each monolayer. 
Local densities in the monolayers are indicated as $\phi^+$ and $\phi^-$ which are defined on the 
bilayer midsurface represented by the dashed line. 
The inter-monolayer friction $w$ also acts at the bilayer midsurface.}
\label{fig:parametrization}
\end{figure}

A vesicle is composed of two opposing layers of lipid molecules whose tails meet at the 
bilayer midsurface.
For a nearly spherical vesicle, this surface can be mathematically described by an angle-dependent radius 
field $r (\theta, \varphi)$.
We assume that the deformation of the vesicle is small enough so that the function $r(\theta, \varphi)$ is a 
continuous single-valued one.
As shown in Fig.~\ref{fig:parametrization}(a), we define a dimensionless deviation in radius $u$ by using the 
radius of a reference sphere $r_0$ as~\cite{Milner87, Faucon89}
\begin{align}
 u (\theta, \varphi ) =  \frac{r( \theta, \varphi )}{r_{0}} - 1.
 \label{eq:definition_u}
\end{align}

A consequence of the membrane bilayer structure is that bending deformation always 
accompanies stretching of one monolayer and compression of the other, as shown in Fig.~\ref{fig:parametrization}(b).
Hence the membrane monolayers are weakly compressible as was found in the 
experiments~\cite{Rawicz00, Sackmann95}.
To take into account this compressibility, the local lipid density on each monolayer is allowed to vary.
Let the number of lipids in the outer and inner monolayers be $N^+$ and $N^-$, respectively.
Throughout this paper, we assign the superscript ``$+$"  to variables associated with the outer monolayer and the fluid outside the vesicle. 
Similarly, we assign the superscript ``$-$" to those associated with the inner monolayer and the fluid enclosed by the vesicle.
Since the spherical vesicle is stable only if the outer monolayer has more lipids than the inner one, 
we impose the restriction $N^+ > N^-$.
We assume that the flip-flop motion of the lipids between the two monolayers does not occur 
within the time scale of experimental observation. 
This means that both $N^+$ and $N^-$ are conserved quantities.
We denote a reference lipid density by $\rho_0 = ( N^+ + N^- ) / ( 2 A_0 )$, where $A_0 = 4\pi r_0^2$
is the total area of the vesicle in the reference state. 
Let $\rho^+ ( \theta, \varphi )$ and $\rho^- ( \theta, \varphi )$ be the variables representing local lipid density 
in each of the monolayers. 
We then define dimensionless local density deviation as
\begin{align}
\phi^\pm (\theta, \varphi) = \frac{\rho^\pm (\theta, \varphi)}{\rho_0} - 1, 
\label{eq:definition_phi_plus_minus}
\end{align}
in each monolayer.
Notice that we have defined the densities on the bilayer midsurface rather than the neutral surface
of a monolayer. 
On the neutral surface, stretching and bending deformations are decoupled by its definition.

Further calculations can be simplified by using the density difference and the density sum defined as
\begin{align}
\phi^\Delta (\theta, \varphi) = \frac{\phi^+ - \phi^-}{2},~~~~~
\phi^\Sigma (\theta, \varphi) = \frac{\phi^+ + \phi^-}{2},
\label{eq:definition_phi_delta_sigma}
\end{align}
respectively. 
At equilibrium, the vesicles take a spherical shape with uniform lipid density in each monolayer. 
In this homogeneous state, the two variables defined above become 
\begin{align}
\phi_0^\Delta = \frac{N^+-N^-}{N^++N^-},~~~~~ 
\phi_0^\Sigma = 0,
\label{eq:ground_state_phi_delta_sigma}
\end{align}
respectively.
Notice that $\phi_0^\Delta>0$ due to the constraint mentioned above. 
The configuration of the vesicle at time $t$ is fully specified using the three dimensionless 
variables, namely, $u ( \theta, \varphi, t)$, $\phi^\Delta ( \theta, \varphi, t)$, and 
$\phi^\Sigma ( \theta, \varphi, t)$.

\subsection{Spherical geometry}

Using the formalism of differential geometry, we briefly introduce the definitions of the relevant 
quantities~\cite{Aris62,Ouyang89}.
Let $\hat{\mathbf{e}}_r$, $\hat{\mathbf{e}}_{\theta}$, and $\hat{\mathbf{e}}_{\varphi}$ be the 
orthonormal unit vectors in the spherical coordinates.
The surface of the vesicle is described using a vector function $\mathbf{r}(\theta, \varphi)$, written as
\begin{align}
\mathbf{r} = r_0[1+u(\theta,\varphi)] \hat{\mathbf{e}}_{r},
\end{align}
where $u$ is defined in Eq.~(\ref{eq:definition_u}).

The two-dimensional (2D) metric tensor of the surface $g_{ij}$ ($i, j = \theta, \varphi$) is defined as
$g_{ij} = \partial_i \mathbf{r} \cdot \partial_j \mathbf{r}$, 
where the notation $\partial_i$ implies a partial derivative with respect to the coordinate $i$.
The inverse of the metric tensor $g^{ij}$ is defined by $g^{ij} g_{jk} = \delta^i_k$,
where $\delta^i_k$ is the Kronecker delta.
Hereafter we use Einstein's summation convention and sum over all repeated indices.
The unit normal vector pointing outward on the surface is given by 
$\hat{\mathbf{n}} = \left( \partial_{\theta} \mathbf{r} \times \partial_{\varphi} \mathbf{r} \right)
/\sqrt{g}$, where $g = \Det g_{ij}$. 
The curvature tensor that quantifies the extrinsic curvature of the surface is further defined as
$L_{ij} = \partial_i \partial_j \mathbf{r} \cdot \hat{\mathbf{n}}$.
Two coordinate-independent scalar curvatures are obtained from $L_{ij}$: 
the mean curvature $H = g^{ij}L_{ij}/2$, and the Gaussian curvature $K = (\Det L_{ij}) / g$.

For a small shape deformation $u$, we express $\sqrt{g}$ and $H$ up to the second order as
\begin{align}
\sqrt{g} \approx r_0^2 \sin \theta \left[ 1 + 2u + u^2 + \frac{(\nabla_\perp u)^2}{2} \right], 
\label{eq:definition_g}
\end{align}
\begin{align}
H \approx -\frac{1}{r_0} \left[ 1 - u - \frac{\nabla_\perp^2 u}{2} + u^2 + u\nabla_\perp^2u \right],
\label{eq:definition_H}
\end{align}
respectively, where we have used 
\begin{align}
&(\nabla_\perp u)^2  = \left(\frac{\partial u}{\partial \theta}\right)^2 + \frac{1}{\sin^2 \theta} \left(\frac{\partial u}{\partial \varphi} \right)^2, 
\end{align}
\begin{align}
&\nabla_{\perp}^2 u = \frac{\partial^2 u}{\partial \theta^2} + \frac{\cos \theta}{\sin \theta} \frac{\partial u}{\partial \theta} + \frac{1}{\sin^2 \theta} \frac{\partial^2 u}{\partial \varphi^2}. \label{eq:definition_laplacian}
\end{align}
Notice that $\nabla_\perp$ is a 2D nabla operator defined on the unit sphere.

\subsection{Vesicle free energy}
\label{subsec:freeenergy}

The thermodynamic free energy of a compressible vesicle includes a curvature elastic energy and a 
stretching elastic energy, and can be written as~\cite{Miao02}
\begin{align}
F  &= \int {\rm d}A \, \Bigg[ \sigma + \frac{\kappa}{2} (2H)^2 + 2\lambda H \phi^{\Delta}  \nonumber \\
&+ k \left[ \left(\phi^\Delta \right)^{2} + \left(\phi^{\Sigma}\right)^{2} \right] \Bigg] +
 \int {\rm d}V \, \Delta P,
\label{eq:freeenergy}
\end{align}
where $\sigma$ is the surface tension, $\kappa$ the bending rigidity, $\lambda$ the 
curvature-density coupling parameter, $k$ the compression modulus, and $\Delta P = P^+ - P^-$
the pressure difference between the outside and the inside of the vesicle in equilibrium.
The first term represents the energy associated with the change in the total area of the vesicle.
The second term describes the energy cost due to the bending deformation~\cite{Helfrich73}. 
The third term represents a coupling between the mean curvature $H$ and 
the local density difference $\phi^{\Delta}$.
The energy associated with compression of the bilayer is described by the fourth term.
The first integral is performed over the whole area $A$ of the vesicle, and hence 
the areal element is given by ${\rm d}A = {\rm d}\theta \, {\rm d}\varphi  \sqrt{g}$.
The volume integral includes the entire volume $V$ of the deformed vesicle.
We assume that the membrane is impermeable, and the fluids inside and outside are incompressible.
This means that the volume of the vesicle is fixed, which is taken into account by the last term in Eq.~(\ref{eq:freeenergy}). 
We note that exactly the same free energy as given by Eq.~(\ref{eq:freeenergy}) was previously employed 
by Miao \textit{et~al.}~\cite{Miao02}.

The physical meaning of the coupling parameter $\lambda$ can be understood such as 
for a lipid bilayer of finite thickness as discussed by Seifert and Langer for planar 
membranes~\cite{Seifert93}.
Since the lipid densities are defined on the bilayer midsurface [see Eq.~(\ref{eq:definition_phi_plus_minus})], 
the local curvature and density are necessarily coupled to each other.
In Ref.~\cite{Seifert93}, they showed that the strength of this coupling is proportional to the distance 
$e$ between the bilayer midsurface and the neutral surface, and the coupling parameter can be interpreted 
as $\lambda = ke$.
Moreover, the bending rigidity $\kappa$ in Eq.~(\ref{eq:freeenergy}) corresponds to the renormalized 
bending rigidity $\tilde{\kappa} = \kappa + 2ke^2$ in Ref.~\cite{Seifert93}.

The parameter $\lambda$ can be controlled either by changing the length of the lipid 
tails, or by attaching a layer of other molecules to the membrane. 
However, this mechanism alone cannot significantly affect the $\lambda$-value.
On the other hand, recent work has shown that an asymmetry in the nature of fluids surrounding 
the bilayer can also influence the curvature-density coupling, even in the case of a single-component 
membrane~\cite{Rozycki15}. 
Moreover, spontaneous curvature can be generally taken into account using a term similar to the 
coupling term in Eq.~(\ref{eq:freeenergy})~\cite{Leibler86,Leibler87}.
In such a case, a strong coupling between the curvature and the membrane internal degree of freedom 
causes a curvature instability.
Since $\lambda$ can be varied by multiple mechanisms, we consider it as a free parameter in our study.

A bilinear coupling between the curvature $H$ and $\phi^\Sigma$ is not allowed because the 
assignment of $\pm$ label is arbitrary.
In accordance with the Gauss-Bonnet theorem, we neglect the Gaussian curvature term $K$ in the 
free energy because it is relevant only when the vesicle topology is allowed to change.

\section{Stability analysis}
\label{sec:statics}

\subsection{Free energy variations}

Here we investigate the stability of the vesicle by considering small deviations of its configuration from 
the reference state.
A spherical vesicle of radius $r_{0}$ with a lipid density difference $\phi_0^\Delta$ and a 
density sum $\phi_0^\Sigma = 0$ is the reference state.
As defined in Eq.~(\ref{eq:definition_u}), the quantity $u$ is a first-order deviation from the reference radius. 
For the density variables, the deviations are given by 
\begin{align}
\delta \phi^{\Delta} = \phi^\Delta - \phi_0^\Delta,~~~~~ 
\delta \phi^{\Sigma} = \phi^{\Sigma},
\label{eq:definition_delta_phi_delta_sigma}  
\end{align}
respectively.
In terms of these deviations, the zeroth order terms of the free energy in Eq.~(\ref{eq:freeenergy}) describe 
the ground state energy of the vesicle
\begin{align}
F^{(0)} &= \frac{4}{3}\pi r_0^3 \Delta P + 4 \pi r_0^2 \sigma + 8 \pi \kappa \nonumber \\
& +  4 \pi r_0^2 k\left(\phi_0^\Delta\right)^2 - 8 \pi r_0 \lambda \phi_0^\Delta.
\label{eq:ground_state_energy}
\end{align}
Next the first order terms in the free energy are
\begin{align}
F^{(1)} &= \left( 2k \left(\phi_0^\Delta\right)^2 - \frac{2 \lambda}{r_0} \phi_0^\Delta + 2 \sigma + \Delta P r_0 \right) \int {\rm d}A_0 \, u \nonumber \\
& + \left(2k \phi_0^\Delta - \frac{2 \lambda}{r_0} \right) \int {\rm d}A_0 \, \delta \phi^{\Delta},
\label{eq:first_order_delta}
\end{align}
where the areal element of the reference sphere is defined here by 
${\rm d}A_0 = r_0^2 \, \sin \theta \, {\rm d}\theta \, {\rm d}\varphi$ and 
should be distinguished from ${\rm d}A$.
The condition for the vesicle to be in mechanical equilibrium is that the first-order variation should vanish, i.e.
$F^{(1)}=0$. 
For this to hold for an arbitrary deviation $u$, we impose the condition
\begin{align}
2 k \left(\phi_0^\Delta \right)^2 - \frac{2 \lambda}{r_0} \phi_0^\Delta + 2 \sigma + \Delta P r_{0} = 0.
\label{eq:capillarity}
\end{align}
This relation is a special case of the more general stability condition of a vesicle when $\lambda$ 
is interpreted as a spontaneous curvature~\cite{Ouyang89}.
In the absence of the first two terms, this is the Young-Laplace equation for surface tension dominated 
interfaces~\cite{Landau87}.
The second term in Eq.~(\ref{eq:first_order_delta}) vanishes because of the assumption that the total 
number of lipids is conserved in each monolayer separately, namely, $\int {\rm d}A_0 \, \phi^\Delta = A_0 \phi_0^\Delta$.

After eliminating $\Delta P$ in the free energy with the use of Eq.~(\ref{eq:capillarity}), we obtain the second-order 
terms in free energy as 
\begin{align}
 F^{(2)} &= \frac{1}{2} \int {\rm d}A_0 \, \Bigg[ \frac{\kappa}{r_0^2} (2 u \nabla_\perp^2 u + (\nabla_\perp^2 u)^2) \nonumber \\
 &- \left[ \sigma + k \left(\phi_0^\Delta\right)^2 - \frac{2\lambda}{r_0} \phi_0^\Delta \right] \left( 2u^2 + u \nabla_\perp^2 u \right) \nonumber \\
 & + \left(8 k \phi_0^\Delta - \frac{4 \lambda}{r_0} \right) \delta \phi^{\Delta}u + \frac{2\lambda}{r_0} \delta \phi^{\Delta} \nabla_\perp^2 u \nonumber \\
  &+ 2k \left[ \left(\delta \phi^{\Delta}\right)^2 + \left(\delta \phi^\Sigma\right)^2 \right] \Bigg].
\end{align}
This second-order free energy will be used to analyze the vesicle stability.

\subsection{Expansions in spherical harmonics}

Since the variables specifying the configuration are all functions of the angular coordinates 
$\theta$ and $\varphi$, we expand them in terms of surface spherical harmonics 
$Y_{nm}(\theta, \varphi)$ as
\begin{align}
u(\theta, \varphi, t) = \sum_{n,m} u_{nm}(t) Y_{nm}(\theta, \varphi),
\label{eq:u_sh_exp}
\end{align}
\begin{align}
\delta \phi^\Delta (\theta, \varphi, t) =  \sum_{n,m}' \phi_{nm}^\Delta (t) Y_{nm}(\theta, \varphi),
\label{eq:phi_delta_sh_exp}
\end{align}
\begin{align}
\delta \phi^{\Sigma} (\theta, \varphi, t) = \sum_{n,m}' \phi_{nm}^\Sigma (t) Y_{nm}(\theta, \varphi).
\label{eq:phi_sigma_sh_exp}
\end{align}
Here, $\sum_{n,m}$ indicates the sum over all spherical harmonic modes　  
$\sum_{n=0}^{\infty} \sum_{m=-n}^{n}$, while 
$\sum_{n,m}'$ in Eqs.~(\ref{eq:phi_delta_sh_exp}) and (\ref{eq:phi_sigma_sh_exp}) denotes 
$\sum_{n=2}^{\infty} \sum_{m=-n}^{n}$.
The $n = 0$ term is omitted in the latter two equations because the total number of lipids is conserved.
The $n=1$ terms are also omitted as they are not coupled to shape changes for small perturbations~\cite{Miao02}.

The spherical harmonics are eigenfunctions of the Laplacian operator on the unit sphere 
$\nabla_\perp^2$, and obey the relation
\begin{align}
\nabla_\perp^2 Y_{nm}(\theta, \varphi) = -n(n+1)Y_{nm}(\theta, \varphi).
\end{align}
By using the spherical harmonics representation, the second-order free energy is written in 
the bilinear form as
\begin{align}
F^{(2)} = \frac{\kappa}{2} \sum_{n,m}' \begin{pmatrix}u_{nm} & \phi_{nm}^\Delta & \phi_{nm}^\Sigma \end{pmatrix} \mathbf{A} \begin{pmatrix}
                            u_{nm} \\[0.8ex]
                            {\phi_{nm}^\Delta} \\[0.8ex]
                            {\phi_{nm}^\Sigma}
  \end{pmatrix}^{\ast},
  \label{eq:freeenergy_bilinear}
\end{align}
where $\ast$ denotes the complex conjugate. 
We have used the orthogonality property of spherical harmonics to obtain Eq.~(\ref{eq:freeenergy_bilinear}).
Since the three configurational variables are real, one has  $u^\ast_{nm} = u_{n, -m}$.
The $3 \times 3$ free energy matrix $\mathbf{A}$ is made dimensionless by dividing all the
components by $\kappa$. 
The symmetric matrix $\mathbf{A}$ is given by
\begin{widetext}
\begin{align}
  \mathbf{A} = \begin{pmatrix}
 (n-1)(n+2) \left[n(n+1) + \sigma' - 2\lambda' \phi_0^\Delta + k' \left(\phi_0^\Delta\right)^{2} \right] \quad & 4k'\phi_0^\Delta-4\lambda' - \lambda'(n-1)(n+2) \quad & 0 \\[1.2ex]
   4k'\phi_0^\Delta-4\lambda' - \lambda'(n-1)(n+2) \quad & 2k' \quad & 0 \\[1.2ex]
   0 \quad & 0 \quad& 2k'
  \end{pmatrix}.
   \label{eq:fem}
\end{align}
\end{widetext}
In the above, the dimensionless parameters are defined by 
\begin{align}
\sigma' = \frac{\sigma r_0^2}{\kappa},~~~~~\lambda' = \frac{\lambda r_0}{\kappa},~~~~~ k' =  \frac{k r_0^2}{\kappa}.
\label{defscaledvariables}
\end{align}

The term proportional to $\left|u_{nm}\right|^2$ in Eq.~(\ref{eq:freeenergy_bilinear}) represents the energy of 
the bending mode. 
Using the equipartition theorem and taking the limits of 
$\lambda'  \to 0$ and $\phi_0^\Delta \to 0$, we obtain the equilibrium fluctuation 
spectrum for the spherical harmonic coefficients as
\begin{align}
\langle \left|u_{nm}\right|^2\rangle = \frac{k_{\rm B}T}{\kappa} \frac{1}{(n-1)(n+2)[n(n+1) + \sigma']},
\end{align}
where $k_{\rm B}$ is the Boltzmann constant and $T$ the temperature~\cite{Safran83}. 
This expression is commonly used to estimate the bending rigidity $\kappa$ and the surface tension 
$\sigma$ of RBCs or vesicles in flicker analysis~\cite{Faucon89, Meleard90}.

\subsection{Stability of the free energy}

The stability of the compressible bilayer vesicle is ensured by the positivity of the eigenvalues of the free energy matrix.
Since $\mathbf{A}$ is a real symmetric matrix, all its eigenvalues are real. 
The matrix $\mathbf{A}$ is further decomposed into one diagonal element $A_{33}=2k'$ 
and $2 \times 2$ sub-matrix $\mathbf{a}$ given by 
\begin{align}
\mathbf{a} = \begin{pmatrix}
A_{11} & A_{12} \\
A_{21} & A_{22}
\end{pmatrix}.
\label{eq:fem_2x2}
\end{align}
The two eigenvalues $\varepsilon_1$ and $\varepsilon_2$ ($\varepsilon_1 < \varepsilon_2$) of the 
matrix $\mathbf{a}$ can be calculated  in terms of its trace and determinant.
Together with $\varepsilon_{3} = 2k'$, they constitute the three eigenvalues of the matrix $\mathbf{A}$. 
The inverse of these three eigenvalues correspond to the ``susceptibilities" of a 
compressible bilayer vesicle, and quantify the proportionality between externally applied forces
and the changes in the three variables $u$, $\phi^\Delta$, and $\phi^\Sigma$.
In other words, the smaller the susceptibility values, the larger the forces required to induce the 
variable changes.

We numerically obtain the two eigenvalues $\varepsilon_{1}$, $\varepsilon_{2}$, using the following 
values for the parameters.
The bending rigidity of membranes has been estimated in several experiments to be about  $\kappa \approx 10^{-19}$~J~\cite{Faucon89,Niggemann95,Rawicz00,Meleard90}.
For diacyl phosphatidylcholine (PC) lipids in their fluid phase, the compression modulus is 
estimated to be $k  \approx 10^{-1}$~J/m$^2$~\cite{Rawicz00,Sackmann95}.
The range of values of the surface tension $\sigma$ changes over four orders of magnitude 
between $10^{-11}$--$10^{-7}$~J/m$^2$.
By considering a typical vesicle radius $r_0 = 10$~$\mu$m, we obtain the parameter values 
$k' = 10^8$ and $\sigma' = 10^{-2}$ [see Eq.~(\ref{defscaledvariables})].

\begin{figure}[tbh]
\centering
\includegraphics[scale=0.48]{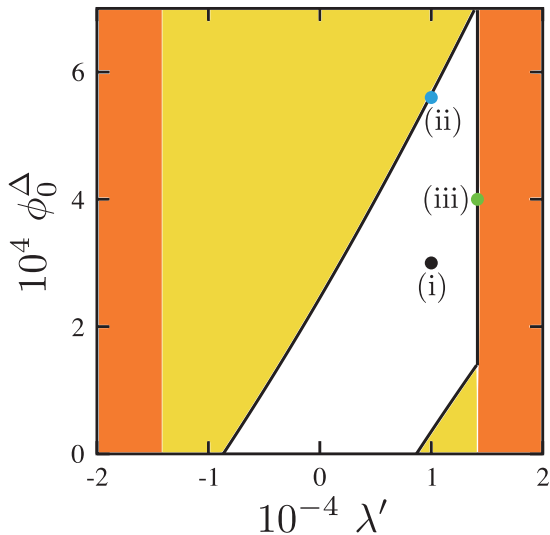}
\caption{Stability diagram in the ($\lambda'$, $\phi_0^\Delta$) plane when $\sigma' = 10^{-2}$
and $k' = 10^{8}$ (see the text).
The vesicle is stable in the white region, while it is unstable in the yellow and orange regions.
In the yellow region, only the $n = 2$ mode is unstable, whereas larger-$n$ modes are unstable
in the orange region.
We investigate the stability and the dynamics represented by the three points inside the stable region:  
(i) $(\lambda', \phi_0^\Delta)=(10^4, 3\times10^{-4})$ (black), 
(ii) $(10^{4}, 5.6\times10^{-4})$ (blue), and 
(iii) $(\sqrt{2} \times 10^{4}, 4 \times 10^{-4})$ (green). 
The case (i) is well inside the stable region, while the cases (ii) and (iii) are very close to the unstable region.}
\label{fig:stability}  
\end{figure}

\begin{figure*}[tbh]
\centering
\includegraphics[scale=0.35]{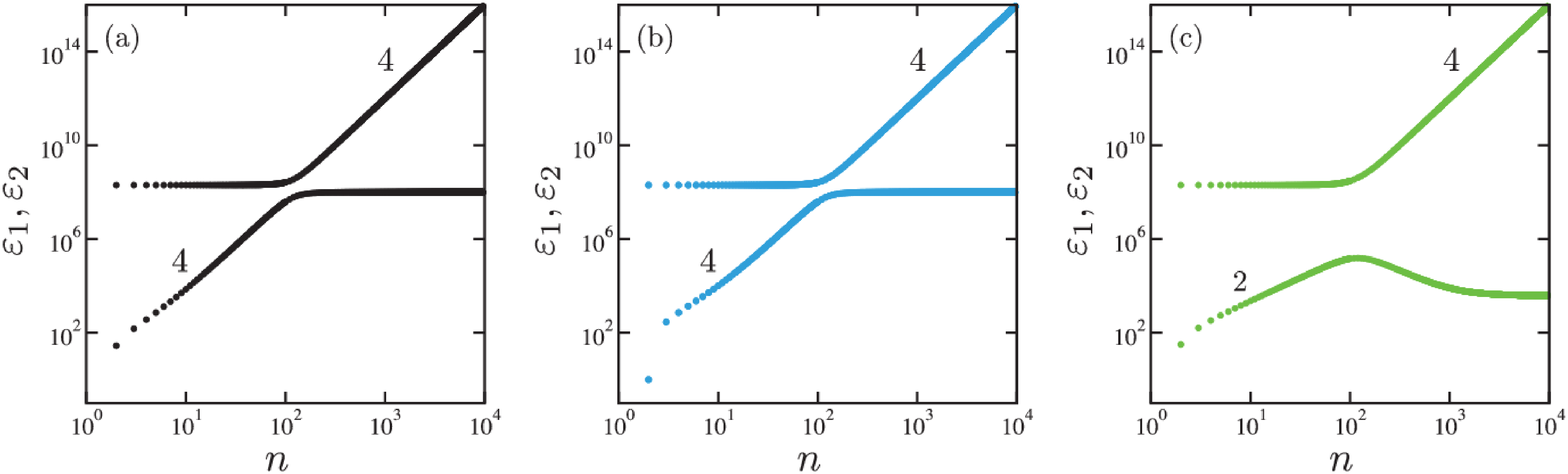}
\caption{The two eigenvalues $\varepsilon_{1}$ and $\varepsilon_{2}$ of the energy matrix 
$\mathbf{A}$ as a function of the spherical harmonic mode $n$ [see Eq.~(\ref{eq:fem})]. 
The chosen parameters are 
(a) $(\lambda', \phi_0^\Delta)=(10^4, 3\times 10^{-4})$ [case (i) in Fig.~\ref{fig:stability}],
(b) $(10^{4}, 5.6\times 10^{-4})$  [case (ii)], and 
(c) $(\sqrt{2} \times 10^{4}, 4 \times 10^{-4})$ [case (iii)].
Other parameter values are $k' = 10^8$ and $\sigma' = 10^{-2}$ as in Fig.~\ref{fig:stability}. 
The numbers in the plots indicate the slopes of different curves.}
\label{fig:eigenvalues}
\end{figure*}

Figure~\ref{fig:stability} shows the stability diagram of the free energy in the parameter space 
($\lambda'$, $\phi_0^\Delta$).
The white region indicates the parameter ranges where all the eigenvalues are positive and the 
vesicle is stable.
Both the yellow and the orange regions represent unstable regions where at least one of the eigenvalues 
is negative.
In the yellow regions, the eigenvalue $\varepsilon_1$ for $n=2$ mode is negative (and hence unstable),
while all other eigenvalues are positive. 
The condition for this instability to occur is approximately obtained by requiring that 
$\Det{\mathbf{a}}$ becomes negative for $n = 2$;
\begin{align}
\left|\phi_0^\Delta  - \frac{3\lambda'}{k'} \right| < \sqrt{\frac{{\lambda'}^2}{k'^2} + \frac{\left(6 + \sigma'\right)}{k'}}.
\end{align}

In the orange regions, the larger $n$-modes become unstable.
The condition for this instability can be also obtained by noticing that the sign of $\varepsilon_{1}$ 
is determined by that of $\Det{\mathbf{a}}$.
Since $\Det{\mathbf{a}} \approx (2k' - \lambda'^2) n^{4}$ for large $n$, the eigenvalue becomes
negative when $|\lambda'| > \sqrt{2k'}$.
Notice that the instability in the orange region is induced only by the parameter $\lambda'$.

In Fig.~\ref{fig:eigenvalues}, we show the two inverse susceptibilities $\varepsilon_1$ and $\varepsilon_2$
as a function of $n$ for three sets of parameters (i), (ii), and (iii) marked in the stable region of 
Fig.~\ref{fig:stability}.
The point (i) is located well inside the stable region, while the points (ii) and (iii) are marginal cases 
with respect to the two instabilities mentioned above. 
We do not plot the third eigenvalue $\varepsilon_3$ as it is merely a constant. 
Figure~\ref{fig:eigenvalues}(a) corresponds to the case (i) where the parameters are 
$\lambda' = 10^4$ and $\phi_0^\Delta = 3\times10^{-4}$.
This value of $\lambda'$ corresponds to a membrane with $e =1$~nm between 
the bilayer midsurface and the neutral surface.
According to Eq.~(\ref{eq:fem}), the energy associated with the pure bending deformation is 
proportional to $n^4$, which is clearly seen in Fig.~\ref{fig:eigenvalues}(a). 
There is also a mode crossing behavior between the $n^4$ bending mode and a constant 
compression mode around $n \approx 100$.

Figure~\ref{fig:eigenvalues}(b) shows the eigenvalues for case (ii) where the parameters
are $\lambda' = 10^4$ and $\phi_0^\Delta = 5.6\times10^{-4}$, which are close to the 
yellow region in Fig.~\ref{fig:stability} where the $n=2$ instability occurs.
The eigenvalue $\varepsilon_1$ for $n=2$ is much smaller than that in Fig.~\ref{fig:eigenvalues}(a).
In the range $2 < n <10$, $\varepsilon_1$ is slightly larger than in (a), while the eigenvalues remain 
almost unchanged for larger $n$.
Figure~\ref{fig:eigenvalues}(c) shows the eigenvalues for the case (iii) where the parameters are 
$\lambda' = \sqrt{2} \times 10^{4}$ and $\phi_0^\Delta = 4 \times 10^{-4}$, which are close to the orange region in 
Fig.~\ref{fig:stability}.
Compared to (a), the $n$-dependence of $\varepsilon_1$ is significantly altered because
we observe $n^2$-dependence for smaller $n$, and $\varepsilon_1$ even decreases for 
larger $n$ values.

The asymptotic value of the smaller eigenvalue is $\varepsilon_1 \approx 2k' - \lambda'^2$
for large $n$. 
In all the three cases (a), (b) and (c), the larger eigenvalue $\varepsilon_2$ remains almost the same. 
It behaves as $\varepsilon_2 \approx 2k'$ for $n \ll 10^2$ and $\varepsilon_2 \approx n^4$ 
for $n \gg 10^2$.

\section{Membrane Hydrodynamics}
\label{sec:dynamics}

\subsection{Onsager's variational principle}

Onsager's variational principle is a fundamental framework to describe the dynamical behavior of 
soft matter~\cite{Doi11}.
It has been applied in a variety of problems including hydrodynamics of liquid crystals, diffusion, and 
sedimentation of colloidal particles~\cite{Doi13}.
In the context of Stokesian hydrodynamics, Onsager's principle states that the viscous forces are balanced 
by the potential forces.
In this approach, a ``Rayleighian" function is defined as the sum of dissipation function
$\Phi$ and the time derivative of the free energy as 
\begin{align}
\mathcal{R} = \Phi + \frac{{\rm d}F}{{\rm d}t}.
\end{align}
The dynamical equations are obtained by extremalizing $\mathcal{R}$ with respect to all the 
dynamical variables.
Any constraint is taken into account by introducing an additional term with a corresponding 
Lagrange multiplier.
A more detailed description on Onsager's variational principle is given in Refs.~\cite{Doi11, Doi13,Fournier15}.

For a compressible bilayer vesicle, we need to obtain the dynamical equations for the three 
variables defined in Sec.~\ref{sec:model}, i.e. $u$, $\phi^\Delta$, and $\phi^\Sigma$.
As we have assumed before,  the fluid medium on either side of the membrane is incompressible.
However, the inside fluid viscosity $\eta^-$ is taken to be different from the outside viscosity $\eta^+$.
The components of the fluid velocity in these regions are represented by $V^\pm_\alpha$, 
and the hydrodynamic pressure fields are denoted by $P^\pm$.
Hereafter, Greek indices are used for components $(r, \theta, \varphi)$ of vectors in three-dimensional 
(3D) space.
The two lipid monolayers are regarded as ideal 2D fluids, and their velocity fields are denoted by $v^\pm_i$.

We take into account two dissipation mechanisms:  the dissipation due to the bulk fluid viscosity,
and the dissipation due to the inter-monolayer friction. 
The dissipation functionals for the outer and inner fluid medium are~\cite{Fournier15},
\begin{align}
\Phi^{\pm} = \int_{\pm} {\rm d}V \, \left( \eta^\pm G^{\sigma \alpha} G^{\rho \beta} D_{\alpha \beta}^\pm D_{\sigma \rho}^\pm - P^\pm G^{\sigma \alpha} \partial_\alpha V_\sigma^\pm \right), 
\label{eq:def_phi_pm}
\end{align}
where $D_{\alpha \beta}^\pm
= ( \partial_\alpha V_\beta^\pm + \partial_\beta V_\alpha^\pm )/2$
is the rate of deformation tensor in the bulk fluid and $G^{\alpha \beta}$ is the inverse of 3D 
(rather than 2D) metric tensor $G_{\alpha \beta}$. 
The $+ $ ($-$) subscript of the integral indicates that it is performed over the volume of the 
outside (inside) bulk fluid.
The dissipation functional due to the inter-monolayer friction is given by 
\begin{align}
\Phi_{w} = \int {\rm d}A_0 \, \frac{w}{2} (v_i^+ - v_i^-)^2, 
\label{eq:def_phi_w}
\end{align}
where $w$ is the friction constant.

Next we discuss the constraints and the hydrodynamic boundary conditions.
Since the monolayer lipid numbers $N^{\pm}$ are conserved, the continuity equation should 
hold in each monolayer separately
\begin{align}
 \frac{\partial \rho^\pm}{\partial t} + g^{ij} \partial_j \left(\rho^\pm v_i^\pm \right) + \frac{\rho^\pm}{2g} \frac{\partial g}{\partial t}  = 0.
\label{eq:general_lipid_mass_cons}
\end{align}
The last term exists due to the curved geometry of the vesicle, and it describes the change in the 
lipid density when the local area element is varied~\cite{Fujitani94, Seki95}.
After the linearization of Eq.~(\ref{eq:general_lipid_mass_cons}), we obtain the following two
equations in terms of small variables 
\begin{align}
\frac{\partial}{\partial t}  \delta \phi^\Delta + \frac{1}{2} (g^{ij}\partial_j v_i^+ - g^{ij}\partial_j v_i^-) = 0, \label{eq:mass_cons_delta} 
\end{align}
\begin{align}
\frac{\partial}{\partial t} \delta \phi^\Sigma + \frac{1}{2} (g^{ij}\partial_j v_i^+ + g^{ij}\partial_j v_i^-) + 2 \frac{\partial u}{\partial t} = 0.
\label{eq:mass_cons_sigma}
\end{align}

The no-slip boundary condition requires that the velocity of the fluid coincide 
with that of the vesicle at the interface. 
Since $u \ll 1$, we approximate that this interface lies at $r = r_{0}$. 
While equating the velocities, the correction due to the finite membrane thickness can be 
neglected because $e/r_{0} \ll u$.
Hence the no-slip condition becomes 
\begin{align}
v_{i}^{\pm} = V_{i}^{\pm},~~~~~
r_{0}\frac{\partial u}{\partial t} = V_{(r)}^{\pm}.
 \label{eq:bc_impenetrable}
\end{align}
Here $V_{(r)}^\pm$ is the physical component of the fluid velocity along the radial direction, and 
should be distinguished from $V_r^\pm$. 
Physical components are the components of a tensor quantity with a correct dimension~\cite{Aris62}. 
In orthogonal coordinate systems, the physical components are related to the covariant components of 
a vector through the relation $h_{\alpha} V_{(\alpha)} = V_{\alpha}$. 
The scale factors $h_{\alpha}$ for the spherical coordinates are 
$(h_{r}, h_{\theta}, h_{\varphi})$ = $(1, r, r\sin \theta)$. 
Additionally, unlike the tensor indices, the quantities in the bracket are not summed over.

Using Eqs.~(\ref{eq:freeenergy}), (\ref{eq:def_phi_pm}), (\ref{eq:def_phi_w}), 
(\ref{eq:mass_cons_delta}), (\ref{eq:mass_cons_sigma}) and (\ref{eq:bc_impenetrable}), 
the total Rayleighian functional for the vesicle and the surrounding fluid is given by
\begin{align}
\mathcal{R} &= \Phi^+ + \Phi^- + \Phi_w + \frac{{\rm d}F}{{\rm d}t} \nonumber \\
& + \sum_{\epsilon = \pm} \int {\rm d}A_0 \, \Bigg[ \mu_i^\epsilon g^{ij}  \left( v_j^\epsilon - V_j^\epsilon \right) + \xi^\epsilon \left( r_0\frac{\partial u}{\partial t} - V_{(r)}^\epsilon \right) \Bigg] \nonumber \\
& + \int {\rm d}A_0 \, \Bigg[\zeta^\Delta \left( \frac{\partial}{\partial t}\delta \phi^\Delta + \frac{1}{2} \left[g^{ij}\partial_jv_i^+ - g^{ij}\partial_jv_i^-\right] \right) \nonumber \\
& + \zeta^{\Sigma} \left(\frac{\partial}{\partial t}\delta \phi^{\Sigma} + \frac{1}{2} \left[g^{ij}\partial_jv_i^+ + g^{ij}\partial_jv_i^-\right] + 2 \frac{\partial u}{\partial t}\right) \Bigg].
\label{eq:totalrayleighian}
\end{align}
Here we have introduced the Lagrange multipliers $\mu_i^\epsilon$ and $\xi^\epsilon$ to take into 
account the boundary conditions in Eq.~(\ref{eq:bc_impenetrable}).
The other Lagrange multipliers $\zeta^\Delta$ and $\zeta^\Sigma$ are used for the continuity equations 
in Eqs.~(\ref{eq:mass_cons_delta}) and (\ref{eq:mass_cons_sigma}), respectively.
The full expression of the Rayleighian is given in the Appendix~\ref{appa}.

\subsection{Basic equations}
\label{sec:basic_equations}

Using the Rayleighian in Eq.~(\ref{eq:totalrayleighian}), we derive the force balance conditions
and hence the dynamical equations by extremalizing it with respect to $V_{\alpha}^{\pm}$, 
$P^\pm$, $\partial_t (\delta \phi^\Delta)$, $\partial_t (\delta \phi^\Sigma)$, 
$\partial_t u $, and $v_i^\pm$, where $\partial_t$ indicates the time derivative.

Extremalizing Eq.~(\ref{eq:totalrayleighian}) with respect to $V_{\alpha}^{\pm}$ yields
\begin{align}
- \eta^\pm G^{\sigma \alpha} \partial_\sigma \partial_\alpha V_\beta^\pm + \partial_\beta P^\pm = 0,
\label{eq:stokes}
\end{align}
which is the Stokes equation for the bulk fluid inside and outside the vesicle.
By extremalizing $\mathcal{R}$ with respect to $P^\pm$, we recover the incompressibility condition of the 
bulk fluid
\begin{align}
G^{\sigma \alpha} \partial_{\alpha} V_{\sigma}^\pm = 0.
\label{eq:incompressibility}
\end{align}

Extremalizing $\mathcal{R}$ with respect to $V_{i}^\pm$ leads to the following equations:
\begin{align}
\mp \eta^\pm \left( \partial_{(r)} V_{(\theta)}^\pm + \frac{1}{r_0}\partial_{(\theta)} V_{(r)}^\pm - \frac{V_{(\theta)}^\pm}{r_0} \right) - \mu_{(\theta)}^\pm = 0 \label{eq:mu_elim}, 
\end{align}
\begin{align}
\mp 2 \eta^\pm \partial_{(r)} V_{(r)}^\pm \pm P^\pm - \xi^\pm = 0. 
\label{eq:xi_elim}
\end{align}
In these equations, we have also included the surface terms which arise from the volume 
integral in $\mathcal{R}$ after performing the integration by parts.
Equation~(\ref{eq:mu_elim}) is obtained by setting $i = \theta$ after the extremalization.
Although a similar equation for the coordinate $\varphi$ can be also obtained,  it is unnecessary for the 
present calculation.
Equations~(\ref{eq:mu_elim}) and (\ref{eq:xi_elim}) are later used to eliminate the Lagrange multipliers 
$\mu_i^\pm$ and $\xi^\pm$.

Extremalizing $\mathcal{R}$ with respect to $\partial_t(\delta \phi^\Delta)$ and 
$\partial_t (\delta \phi^\Sigma)$, we obtain
\begin{align}
\zeta^\Delta + 2k\delta \phi^\Delta + \left(4k\phi_0^\Delta - \frac{4 \lambda}{r_0} \right) u 
+ \frac{\lambda}{r_0}\left(2u + \nabla_\perp^2 u\right) = 0,
\label{eq:zeta_delta_elim} 
\end{align}
\begin{align}
\zeta^{\Sigma} + 2k\delta \phi^{\Sigma} = 0,
\label{eq:zeta_sigma_elim}
\end{align}
respectively, which are also used to eliminate $\zeta^{\Delta}$ and $\zeta^{\Sigma}$. 
The force balance condition along the radial direction is obtained by extremalizing $\mathcal{R}$ 
with respect to $\partial_t u$ 
\begin{align}
  & r_0 (\xi^+ + \xi^-) + 2 \zeta^\Sigma - k \left(\phi_0^\Delta\right)^{2} \left(2u + \nabla_\perp^2 u \right) + 4k \phi_0^\Delta \delta \phi^{\Delta} \nonumber \\
  &  - \frac{2 \lambda}{r_0} \delta \phi^{\Delta} + \frac{\lambda}{r_0} \nabla_\perp^2 \delta \phi^\Delta  + \frac{2 \lambda}{r_0} \phi_0^\Delta \nabla_\perp^2 u + \frac{4 \lambda}{r_0} \phi_0^\Delta u\nonumber \\
  &  + \frac{\kappa}{r_0^2} \left( 2 \nabla_\perp^2 u + \nabla_\perp^4 u \right) - \sigma (2u + \nabla_\perp^2u)  = 0.
  \label{eq:ldot_min}
\end{align}

Finally, we extremalize $\mathcal{R}$ with respect to $v_{i}^\pm$ and obtain the lateral force balance 
on each of the monolayers as 
\begin{align}
\mu_i^\pm + w \left( v_i^\pm - v_i^{\mp} \right) \mp \frac{1}{2} \partial_i\left(\zeta^{\Delta} \pm \zeta^{\Sigma} \right) = 0.
\label{eq:vi_min}
\end{align}

\subsection{Solutions of hydrodynamic equations}

Next we use the Lamb's solution for Stokes equation in Eq.~(\ref{eq:stokes}) in the spherical 
coordinates~\cite{Lamb75, Happel73, Seki95}. 
It is represented in terms of the pressure field $P^\pm(\mathbf{r},t)$ and the 
newly introduced function $\psi^\pm(\mathbf{r}, t)$ that is the solution of the homogeneous 
Stokes equation $G^{\sigma \alpha} \partial_{\sigma} \partial_{\alpha} V_{\beta}^{\pm} = 0$. 
We expand these functions in terms of solid spherical harmonics as below
\begin{align}
\psi^\pm (\mathbf{r},t) = \sum_{n,m}' \psi_{nm}^\pm(t) \left(\frac{r}{r_0}\right)^n Y_{nm}(\theta, \varphi),
\end{align}
\begin{align}
P^\pm (\mathbf{r},t) = \sum_{n,m}' P_{nm}^\pm(t) \left(\frac{r}{r_0}\right)^n Y_{nm}(\theta, \varphi).
\end{align}
Then the Lamb's solution can be written as 
\begin{align}
\mathbf{V}^{+} (\mathbf{r}) &= \sum_{n,m}' \Bigg( \psi_{nm}^{+} \nabla - \frac{(n-2)}{2\eta^{+} n (2n - 1)} P_{nm}^{+} r^{2} \nabla \nonumber \\
& + \frac{(n+1)}{\eta^{+}n(2n - 1)} P_{nm}^{+} \mathbf{r} \Bigg)  \left( \frac{r_0}{r} \right)^{n+1} Y_{nm}(\theta, \varphi),
\end{align}
\begin{align}
  \mathbf{V}^{-} (\mathbf{r}) &= \sum_{n,m}' \Bigg( \psi_{nm}^{-} \nabla + \frac{(n+3)}{2\eta^{-} (n + 1) (2n + 3)} P_{nm}^{-} r^{2} \nabla \nonumber \\
&- \frac{n}{\eta^{-}(n + 1)(2n + 3)} P_{nm}^{-} \mathbf{r} \Bigg) \left( \frac{r}{r_{0}} \right)^{n} Y_{nm}(\theta, \varphi).
\end{align}

The velocity field of the lipid monolayer $\mathbf{v}^\pm$ is expanded in terms of vector spherical harmonics.
Here, only the component along $\nabla Y_{nm}$ is relevant for our calculation, and it is denoted as 
$v_{nm}^\pm$. 
Using the boundary conditions in Eq.~(\ref{eq:bc_impenetrable}), we equate the components along $\nabla Y_{nm}$ to obtain
\begin{align}
v_{nm}^{+} = \frac{1}{r_{0}} \left( \psi_{nm}^{+} - \frac{(n-2)}{2\eta^{+} n (2n - 1)} P_{nm}^{+} r_{0}^{2} \right),
\end{align}
\begin{align}
v_{nm}^{-} = \frac{1}{r_{0}} \left( \psi_{nm}^{-} + \frac{(n+3)}{2\eta^{-} (n + 1) (2n + 3)} P_{nm}^{-} r_{0}^{2} \right).
\end{align}
These equations are used in the force balance conditions Eqs.~(\ref{eq:ldot_min}) and (\ref{eq:vi_min}) 
to derive the final dynamical equations as derived in the Appendix~\ref{appb}.

\subsection{Dynamical equations}

As shown in the Appendix~\ref{appb}, the dynamical equation for the variable $u$ 
is given by
\begin{align}
\frac{\partial}{\partial t} u_{nm} = \frac{n}{r_0^2} \psi_{nm}^- + \frac{n}{2\eta^-(2n + 3)} P_{nm}^{-},
\label{eq:unm_dynamical}
\end{align} 
[see Eq.~(\ref{eq:impenetrable_sh})].
The dynamical equations for the density variables $\phi^{\Delta}$ and $\phi^{\Sigma}$ are derived 
from Eqs.~(\ref{eq:mass_cons_delta}) and (\ref{eq:mass_cons_sigma}) as
\begin{align}
\frac{\partial}{\partial t}\delta \phi_{nm}^\Delta &= \frac{n(n+1)}{2r_0^2} \Bigg( \psi_{nm}^+ - \frac{(n-2)r_0^2}{2 \eta^+ n (2n-1)}P_{nm}^+ \nonumber \\
& - \psi_{nm}^- - \frac{(n+3)r_0^2}{2 \eta^-(n+1)(2n+3)}P_{nm}^- \Bigg),
\label{eq:mass_cons_plus_sh}
\end{align}
\begin{align}
\frac{\partial}{\partial t}\delta \phi_{nm}^\Sigma &= \frac{n(n+1)}{2r_0^2} \psi_{nm}^+ - \frac{(n+1)(n-2)}{4 \eta^+(2n-1)}P_{nm}^+  \nonumber \\
&+ \left( \frac{n(n+1)}{2} - 2n \right) \frac{\psi_{nm}^-}{r_0^2} \nonumber \\
&+ \left( \frac{n(n+3)}{4 \eta^-(2n+3)} - \frac{n}{\eta^-(2n+3)} \right) P_{nm}^-,
\label{eq:mass_cons_minus_sh}
\end{align}
respectively.

By using the solutions for $\psi_{nm}^\pm$ and $P_{nm}^\pm$, as obtained in the Appendix~\ref{appb},
the final dynamical equations are expressed in the matrix form as 
\begin{align}
  \frac{\partial}{\partial t}
\begin{pmatrix}
u_{nm} \\[0.8ex]
\phi_{nm}^\Delta \\[0.8ex]
\phi_{nm}^\Sigma 
\end{pmatrix} = - \frac{1}{\tau} \mathbf{B} \begin{pmatrix}
								u_{nm} \\[0.8ex]
								\phi_{nm}^\Delta \\[0.8ex]
								\phi_{nm}^\Sigma
								\end{pmatrix}.
\label{eq:dynamical_eqs}								
\end{align}
Here the $3 \times 3$ matrix $\mathbf{B}$ is made dimensionless by using the characteristic time scale
\begin{align}
\tau = \frac{\eta^+ r_0^3}{\kappa}.
\label{eq:tau}
\end{align}
We also define the viscosity contrast $E$ as the ratio between the viscosities inside and outside of the vesicle
\begin{align}
E = \frac{\eta^-}{\eta^+}.
\end{align}
We further introduce a dimensionless friction coefficient given by 
\begin{align}
w' = \frac{wr_0}{\eta^+}.
\end{align}

The dynamical matrix $\mathbf{B}$ in Eq.~(\ref{eq:dynamical_eqs}) is expressed as a 
product of two symmetric matrices
\begin{align}
\mathbf{B} = \mathbf{C} \cdot \mathbf{A},
\label{eq:full_dynamical_matrix}
\end{align}
where $\mathbf{A}$ was defined before in Eq.~(\ref{eq:fem}) as a free energy matrix. 
On the other hand, the matrix $\mathbf{C}$ depends only on $E$ and $w'$ which characterize 
the dissipation of the whole system. 
Since the full analytical expression for the matrix $\mathbf{C}$ is lengthy, we present its 
each component in the Appendix~\ref{appc} which provides the main result of our analytic 
calculation.

\begin{figure}[tbh]
\centering
\includegraphics[scale=0.48]{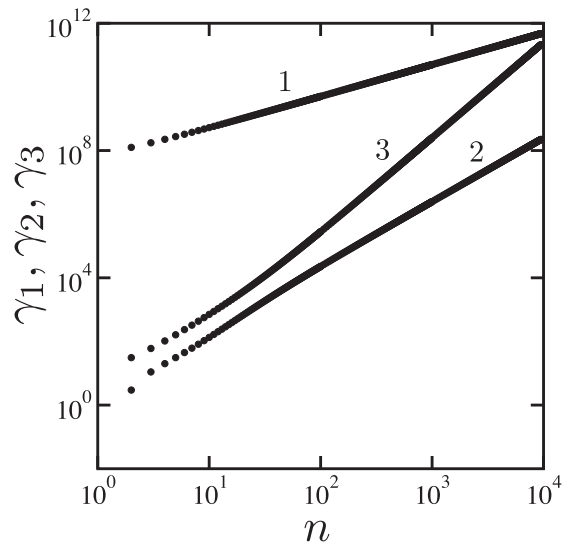}
\caption{The three relaxation rates $\gamma_1$,  $\gamma_2$, and  $\gamma_3$ 
as a function of the spherical harmonic mode $n$ obtained from the dynamical 
matrix $\mathbf{B}$ [see Eq.~(\ref{eq:full_dynamical_matrix})].
The chosen parameter values are 
$(\lambda', \phi_0^\Delta)=(10^4, 3\times10^{-4})$ (case (i) in Fig.~\ref{fig:stability}), 
$k' = 10^8$, $\sigma' = 10^{-2}$, and $w' = 10^7$. 
The fluid viscosities are taken to be symmetric between inside and outside of the vesicle, i.e. 
$E =\eta^-/\eta^+= 1$.
All the relaxation rates are scaled by the characteristic time $\tau = \eta^{+} r_{0}^{3} / \kappa$.
The numbers indicate the slopes of different curves.
}
\label{fig:relaxation}
\end{figure}

\section{Relaxation modes}
\label{sec:results}

\subsection{Three relaxation modes}

The coupled equations for the relaxation dynamics of a weakly compressible bilayer vesicle using the three variables 
$u$, $\phi^\Delta$, and $\phi^\Sigma$ are obtained in Eq.~(\ref{eq:dynamical_eqs}).
The three eigenvalues of the dynamical matrix $\mathbf{B}$, denoted as $\gamma_1 < \gamma_2 < \gamma_3$, 
give the relaxation rates as a function of $n$.
The presence of the off-diagonal elements in $\mathbf{B}$ indicates that these variables are coupled to each other. 
This is a fundamental and unique feature of a spherical vesicle, because the mode associated with 
$\phi^\Sigma$ is always decoupled from the other two modes for planar bilayers~\cite{Seifert93}.

As for the dynamical parameters, the inter-monolayer friction is estimated to be about 
$w \approx 10^9$~N$\cdot$s/m$^3$ according to the vesicle fluctuation analysis~\cite{Pott02}.
The viscosity of the external fluid is set to be $\eta^+ = 10^{-3}$~Pa$\cdot$s for water.
Hence, in addition to the dimensionless static parameters given in Sec.~\ref{sec:statics}, 
we use here $w' = 10^7$.
With these parameter values, the characteristic time scale in Eq.~(\ref{eq:tau}) becomes $\tau = 10$~s.
We first consider the equal viscosity case $E=1$ and numerically calculate the three eigenvalues 
in Fig.~\ref{fig:relaxation} when $\lambda' = 10^4$ and $\phi_0^\Delta = 3\times10^{-4}$
(case (i) in Fig.~\ref{fig:stability}). 
Notice that all the relaxation rates are made dimensionless here by using $\tau$.
In order to observe the mode crossing behavior, we have computed 
up to $n = 10^4$ which corresponds to few nm wavelength for a vesicle of size $r_0=10$~$\mu$m.

From Fig.~\ref{fig:relaxation}, we find that the two eigenvalues $\gamma_1$ and $\gamma_2$ 
are asymptotically proportional to $n^2$ and $n^3$, respectively, for large $n$. 
In the later subsection, we show that the $n^2$-dependence predominantly corresponds to the 
slipping relaxation mode, while the $n^3$-dependence is associated with the bending relaxation mode.
At around $n \approx 20$, these two modes start to overlap each other, which is a mode crossing 
behavior.
The presence of the slipping mode and the cross-over between the two relaxation modes 
were first considered by Seifert \textit{et~al.}\ for planar bilayers~\cite{Seifert93}.
Indeed neutron spin echo and flicker spectroscopy studies of GUVs have confirmed that 
large-$n$ relaxation is dominated by the slipping mode~\cite{Mell15, Rodriguez-Garcia09}.

The three relaxation rates for $n=2$ mode are estimated to be $0.296$~s$^{-1}$, $3.097$~s$^{-1}$, 
and $1.257 \times 10^7$~s$^{-1}$.
The smallest relaxation rate obtained here is close to that measured using flicker analysis~\cite{Meleard90}.
We also notice that the largest eigenvalue $\gamma_3$ is proportional to $n$ throughout the range 
of $n$ studied, and corresponds to the relaxation of the total density $\phi^\Sigma$~\cite{Seifert93, Miao02}.
This mode is much (about eight orders of magnitude) faster than the other two slower ones.

\subsection{Elimination of the fastest mode}

Since the relaxation of the total density is much faster compared to the other two slower modes, 
we are allowed to eliminate it from the relaxation equations. 
Hence we assume that the variable $\phi^\Sigma$ equilibrates immediately, and
set $\partial \phi^\Sigma / \partial t \approx 0$.
Within this approximation, we can eliminate $\phi^\Sigma$ from the coupled dynamical equations
Eq.~(\ref{eq:dynamical_eqs}) to obtain 
\begin{align}
  \frac{\partial}{\partial t}
\begin{pmatrix}
u_{nm} \\[0.8ex]
\phi_{nm}^\Delta
\end{pmatrix} = - \frac{1}{\tau} \mathbf{b} \begin{pmatrix}
     u_{nm} \\[0.8ex]
     \phi_{nm}^\Delta
\end{pmatrix}.
\label{eq:reduced_dynamical_matrix}
\end{align}
Here the $2\times2$ matrix $\mathbf{b}$ can be again written as a product of two symmetric matrices as 
\begin{align}
\mathbf{b} = \mathbf{c} \cdot \mathbf{a},
\label{dynamicalmatrxb}
\end{align}
where $\mathbf{a}$ was defined in Eq.~(\ref{eq:fem_2x2}), and  the matrix $\mathbf{c}$ takes the form
\begin{widetext}
\begin{align}
\mathbf{c} = \frac{n(n+1)}{d}\begin{pmatrix}
            4w' + (2n+1)(1+E) \quad & -  (n+2) + (n-1)E  \\[1.2ex]
            -  (n+2) + (n-1)E  \quad & (n+2)(2n^{2}-n+2) + (n-1)(2n^{2}+5n+5)E 
\end{pmatrix}.
\label{eq:matrix_c}
\end{align}
\end{widetext}
In the above, the common denominator $d$ is given by
\begin{align}
d &= 4 w' \left[(n-1)(2n^2+5n+5)E + (n+2)(2n^2-n+2) \right] \nonumber \\
&+ 2(n-1)(n+1)(2n^2+4n+3)E^2 \nonumber \\
& + \left[8n^2(n+1)^{2}-5\right]E + 2n(n+2)(2n^2+1).
\label{eq:d_hat}
\end{align}

\begin{figure}[tbh]
\centering
\includegraphics[scale=0.48]{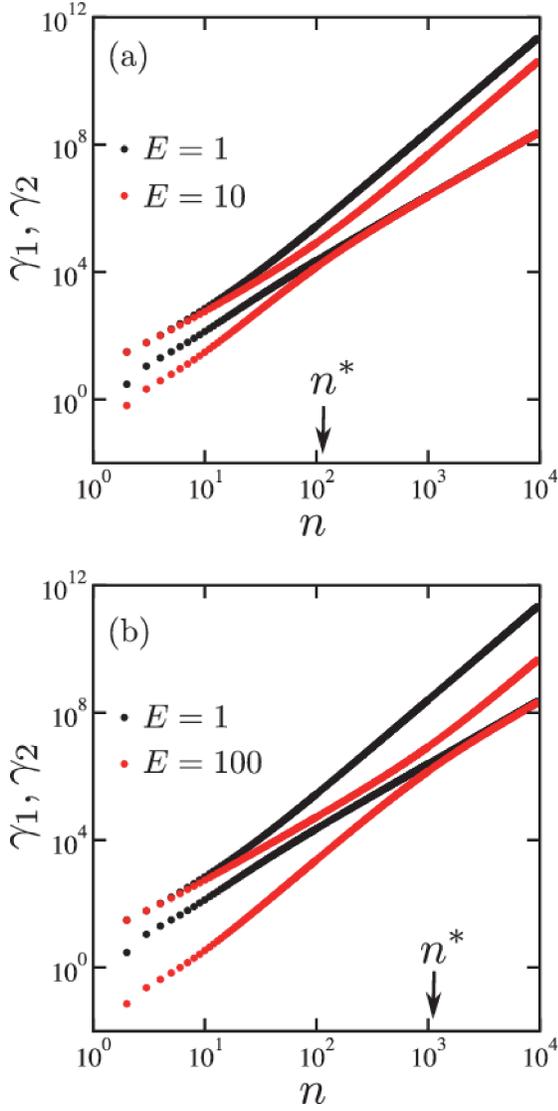}
\caption{The two slower relaxation rates $\gamma_1$ and $\gamma_2$ as a function of the spherical 
harmonic mode $n$ obtained from the dynamical matrix $\mathbf{b}$ [see Eq.~(\ref{dynamicalmatrxb})]. 
The chosen parameter values are $(\lambda', \phi_0^\Delta)=(10^4, 3\times10^{-4})$ (case (i) 
in Fig.~\ref{fig:stability}), $k' = 10^8$, $\sigma' = 10^{-2}$, and $w' = 10^7$. 
(a) The comparison between $E=\eta^-/\eta^+=1$ (black) and $E=10$ (red).
(b) The comparison between $E=1$ (black) and $E=100$ (red).
All the relaxation rates are scaled by the characteristic time $\tau = \eta^{+} r_{0}^{3} / \kappa$.
Arrows indicate the cross-over mode $n^{\ast}$ given by Eq.~(\ref{crossovermode}).
}
\label{fig:viscosity_contrast} 
\end{figure}

In Fig.~\ref{fig:viscosity_contrast}, we plot the two eigenvalues of the matrix $\mathbf{b}$ 
(black symbols) for the same parameter values used in Fig.~\ref{fig:relaxation}.
The two relaxation rates obtained here completely coincide with $\gamma_1$ and $\gamma_2$ 
in Fig.~\ref{fig:relaxation}.
Such a coincidence justifies our approximation to eliminate the fastest relaxation mode associated 
with $\phi^\Sigma$.

\begin{figure}[tbh]
\centering
\includegraphics[scale=0.48]{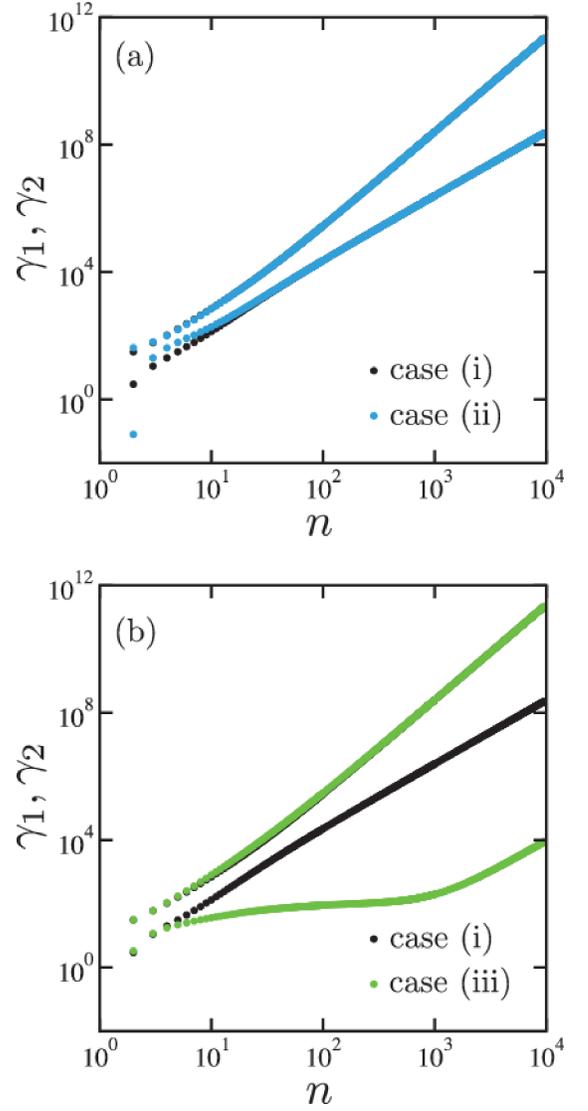}
\caption{The two slower relaxation rates $\gamma_1$ and $\gamma_2$ as a function of the spherical 
harmonic mode $n$ obtained from the dynamical matrix $\mathbf{b}$ [see Eq.~(\ref{dynamicalmatrxb})]. 
The chosen parameter values are $k' = 10^8$, $\sigma' = 10^{-2}$, $w' = 10^7$ and 
$E=\eta^-/\eta^+=1$. 
(a) The comparison between the stable case (i) (black) and the marginal case (ii) (blue) in 
Fig.~\ref{fig:stability}.
(b) The comparison between the stable case (i) (black) and the marginal case (iii) (green) in 
Fig.~\ref{fig:stability}.
All the relaxation rates are scaled by the characteristic time $\tau = \eta^{+} r_{0}^{3} / \kappa$.
}
\label{fig:unstable_dynamics} 
\end{figure}

\subsection{Effects of different viscosities}
\label{subsec:viscosity}

In this subsection, we discuss the effect of viscosity contrast $E=\eta^-/\eta^+$ on the relaxation 
dynamics of a compressible bilayer vesicle.
Figure~\ref{fig:viscosity_contrast}(a) shows the relaxation rates when the inner viscosity is 10 
times larger than that of the outside, i.e. $E = 10$ (red symbols), while all the other parameters are 
same as the case of $E=1$ (black symbols). 
The mode crossing behavior between the bending and the slipping modes at around $n \approx 200$
is more apparent in the higher viscosity contrast case.
When $E=10$, the bending mode becomes slower than that of $E=1$ case, while the slipping 
mode remains almost unaffected.
The effect of viscosity contrast is more pronounced in the case of $E = 100$ (red symbols) as shown 
in Fig.~\ref{fig:viscosity_contrast}(b).
Here the slowest relaxation mode is dominated by the bending one up to $n \approx 2000$.
Such a relaxation behavior is significantly different from the equal viscosity case, where the 
slipping mode dominates most of the relevant $n$-range.
In Fig.~\ref{fig:viscosity_contrast}(b), we find that the relaxation rate 
$\gamma_1$ for small $n$ values differs by two orders of magnitude between the cases 
$E = 1$ and $100$.

As shown in Fig.~\ref{fig:viscosity_contrast}, the viscosity contrast shifts the mode crossing point 
between the slipping and the bending modes.
To better understand this effect, we approximate one of the eigenvalues of the matrix 
$\mathbf{b}=\mathbf{c} \cdot \mathbf{a}$ as follows. 
Keeping only the most dominant term, one can approximate Eqs.~(\ref{eq:matrix_c}), (\ref{eq:d_hat}),
and (\ref{eq:fem_2x2}) as 
\begin{align}
\mathbf{c} \approx \frac{n(n+1)}{d}\begin{pmatrix}
            4 w' & 0 \\
            0 & 2n^{3}(1+E)
            \end{pmatrix},
\end{align}
\begin{align}
d \approx 8 w' n^{3} (1+E),
\end{align}
\begin{align}
\mathbf{a} \approx \begin{pmatrix}
            n^{4} & -\lambda' n^{2} \\
            -\lambda' n^{2} & 2k'
            \end{pmatrix}.
\end{align}
Then the larger eigenvalue of $\mathbf{b}$ is approximated as 
\begin{align}
  \gamma_2 & \approx \frac{w' n^{3} + k'n^{2}(1+E)}{2w' (1+E)} \nonumber \\
       & \approx \begin{cases}
         \dfrac{k' n^2}{2 w'} & n \ll n^\ast \\[1.5ex]
         \dfrac{n^3}{2(1+E)} & n \gg n^\ast.
         \end{cases}
 \label{crossover}
\end{align}
where
\begin{align}
n^\ast = \frac{k'}{w'} (1+E)
\label{crossovermode}
\end{align}
is the cross-over mode and is linearly proportional to the viscosity contrast $E$.

The above limiting expressions clearly show that the $n^2$-dependence corresponds to the slipping mode,
while the $n^3$-dependence reflects the bending mode.
Moreover, only the bending mode is affected by the viscosity contrast $E$ as it appears 
only for $n \gg n^\ast$ in Eq.~(\ref{crossover}).
The above limiting expressions of the relaxation rates coincide with those for a planar 
bilayer~\cite{Seifert93} when $E=1$ and  $n/r_0$ is regarded as a wavenumber 
in the limit of $r_0 \to \infty$.

\subsection{Dynamics close to the unstable regions}

As we have discussed in Sec.~\ref{sec:statics}, the two distinct instabilities are identified in 
the $(\lambda',\phi_0^\Delta)$ parameter plane.
Here we focus on the relaxation dynamics of a vesicle when the parameters are close to these 
unstable regions as marked by the points (ii) and (iii) in Fig.~\ref{fig:stability}. 
In Fig.~\ref{fig:unstable_dynamics}(a), we plot the two relaxation rates from the matrix $\mathbf{b}$
as a function of $n$ when the parameter values are that of point (ii), i.e. 
$\lambda' = 10^4$ and $\phi_0^\Delta = 5.6\times10^{-4}$.
Compared to the case (i), the relaxation rate $\gamma_1$ with $n = 2$ is significantly reduced. 
The relaxation rate $\gamma_1$ in the range $2 < n < 10$ is slightly larger than that of case (i), 
and they are almost the same for larger $n$.
The relaxation rates $\gamma_2$ does not vary between the cases (i) and (ii), except for a small 
increase in the case of (ii) when $n = 2$.

In Fig.~\ref{fig:unstable_dynamics}(b), we plot the two relaxation rates from the matrix $\mathbf{b}$
when the parameter values are that of the case (iii), i.e. $\lambda' = \sqrt{2} \times 10^4$ and $\phi_0^\Delta = 4\times10^{-4}$.
As compared to the case of point (i), the large-$n$ relaxation of the slipping mode is slowed down 
by more than four orders of magnitude. 
Moreover, the slipping mode is almost independent of $n$ in the intermediate-$n$ range
($150 < n <500$) for the case (iii).
The relaxation rate $\gamma_2$ is not significantly different between the cases (i) and (iii).

\subsection{Limiting expressions}

Here we make a connection of our work to that by Seki \textit{et~al.}~\cite{Seki95}, where they 
investigated the relaxation dynamics of a compressible vesicle without considering a bilayer structure.
In the present work, this corresponds to the case in which the friction between the monolayers is large 
enough such that the two monolayers move together as a single layer, i.e. $w' \to \infty$.
Moreover, we also take the limit of $\lambda' \to 0$, $\phi_0^\Delta \to 0$ in order to recover the 
previous result.

In this limit, all the coefficients of $\phi^{\Delta}$ in the dynamical equation in Eq.~(\ref{eq:dynamical_eqs}) 
vanish resulting in a reduced $2\times2$ dynamical matrix
\begin{equation}
\frac{\partial}{\partial t}\begin{pmatrix}
u_{nm} \\[0.8ex]
\phi_{nm}^\Sigma 
\end{pmatrix} = - \frac{1}{\tau} \tilde{\mathbf{b}} \begin{pmatrix}
								u_{nm} \\[0.8ex]
								\phi_{nm}^\Sigma
								\end{pmatrix}.
\end{equation}
Again, the matrix $\tilde{\mathbf{b}}$ can be expressed as a product of two symmetric matrices  
\begin{align}
\tilde{\mathbf{b}} = \tilde{\mathbf{c}} \cdot \tilde{\mathbf{a}}.
\end{align}
Here the matrix $\tilde{\mathbf{c}}$ is 
\begin{widetext}
\begin{align}
\tilde{\mathbf{c}} = \frac{n(n+1)}{\tilde{d}}\begin{pmatrix}
            (2n+1)(1+E) \quad & - (n+2) + (n-1)E \\[1.2ex]
            - (n+2) + (n-1)E \quad & (n+2)(2n^{2}-n+2) + (n-1)(2n^{2}+5n+5)E 
            \end{pmatrix},
\end{align}
\end{widetext}
with a common denominator 
\begin{align}
\tilde{d} &= \left[2(n-1)(n+1)E + 2 n^2 + 1\right] \nonumber \\
&\times \left[(2 n^2 + 4n + 3) E + 2n (n+2)\right].
\end{align}
On the other hand, the matrix $\tilde{\mathbf{a}}$ is given by
\begin{align}
  \mathbf{\tilde{a}} = \begin{pmatrix}
    (n-1)(n+2)[n(n+1) + \sigma'] & 0 \\[1.2ex]
    0 & 2k'
  \end{pmatrix}.
\end{align}

Following the work of Seki \textit{et~al.}~\cite{Seki95}, we obtain two limiting expressions for the 
relaxation rates.
When the membrane compressibility is large compared to the bending rigidity, i.e. $n(n+1) + \sigma' \ll k'$, 
we get
\begin{align}
\gamma_n = \frac{[n (n+1) + \sigma'](n-1)n(n+1)(n+2)}{(n-1)(2n^{2}+5n+5)E + (n+2)(2n^{2}-n+2)}.
\label{eq:limit1}
\end{align}
If we further set $E = 1$, this expression reduces to that obtained first by Milner and 
Safran~\cite{Milner87}. 
When the bending rigidity dominates over the compressibility $k' \ll n(n+1) + \sigma'$, we obtain
\begin{align}
\gamma_n = \frac{2k'n(n+1)}{(2n+1)(1+E)}.
\label{eq:limit2}
\end{align}
The limiting expressions Eqs.~(\ref{eq:limit1}) and (\ref{eq:limit2}) exactly coincide with 
Eqs.~(D.6) and (D.7) in Ref.~\cite{Seki95}, respectively.

\section{Summary and discussions}
\label{sec:discussions}

In this paper, we have investigated the statics and dynamics of a weakly compressible bilayer vesicle.
First we have calculated a free energy matrix in terms of linear perturbations in local curvature 
and local densities.
Calculating the eigenvalues of the free energy matrix, we performed the stability analysis by varying 
the curvature-density coupling parameter $\lambda'$, and the lipid density difference between the 
two monolayers $\phi_0^\Delta$.
As a result, we identified two different instabilities: one affecting the small-$n$ modes and the other 
influencing the large-$n$ modes.

Onsager's variational principle offers an universal framework for obtaining the dynamical equations 
for soft matter, and it has been used to derive the dynamical equations of a compressible bilayer 
vesicle.
The eigen values of the dynamical matrix $\mathbf{B}$ in Eq.~(\ref{eq:full_dynamical_matrix}) gives 
the relaxation rates for a vesicle whose inside and outside fluids are characterized by different viscosities.
The three relaxation modes are coupled to each other as a consequence of the bilayer and the spherical
structure of the vesicle. 
Assuming that one of the relaxation modes is much faster than the other two, we derived the dynamical equation for the slower modes in Eq.~(\ref{eq:reduced_dynamical_matrix}). 
We focused on the effect of viscosity contrast $E=\eta^-/\eta^+$ on the relaxation rates, and found that 
it linearly shifts the cross-over $n$-mode between the bending and the slipping relaxations
[see Eq.~(\ref{crossovermode})].
As $E$ is increased, the relaxation rate of the bending mode decreases, while that of the slipping mode
remains almost unaffected.
For parameter values close to the unstable region, some of the relaxation modes are dramatically reduced.
For example, as they approach the region of large-$n$ instability, we find an unusual behavior of the 
relaxation rate as shown in Fig.~\ref{fig:unstable_dynamics}(b).

Although the viscosity contrast in vesicles and RBCs has been discussed in some 
experiments~\cite{Fujiwara14, Turlier16}, we have derived here the exact relaxation rates of a 
compressible bilayer vesicle for arbitrary $E$-values.
In experiments, a commonly encountered case is where the inside viscosity is slightly larger than 
the outside $E > 1$, for which the bending mode is slowed down and cross-over mode $n^{\ast}$ 
becomes larger.
For even larger viscosity contrast ($E \gg 100$), the relaxation is entirely dominated by the bending mode.
For $E < 1$, on the other hand,  $n^{\ast}$ does not depend on $E$ and the relaxation rate of the 
bending mode is slightly increased.
Our result clearly shows that the viscosity contrast significantly affects the dynamical behavior of a vesicle.

In the present work, the vesicle stability was analyzed in terms of the curvature-density coupling parameter 
$\lambda'$, and the lipid density difference between the monolayers $\phi_0^\Delta$.
Whereas the large-$n$ instability is induced only by $\lambda'$, the small-$n$ instability can be 
triggered by changing either $\lambda'$ or $\phi_0^\Delta$.
Since the large-$n$ instability corresponds to small wavelength deformations, it leads to the stabilization 
of small buds or tube-like deformations.
For parameter ranges close to the yellow region in Fig.~\ref{fig:stability}, we predict that the small-$n$ 
modes slow down significantly. 
Although it would be experimentally challenging, the control of the number of lipids in each monolayers
enables the change of the parameter $\phi_0^\Delta$.

Here we shall briefly mention the relation of our result to the previous theoretical works. 
Extending the work by Schneider \textit{et~al.}~\cite{Schneider84}, Milner and Safran derived the 
bending relaxation rate in vesicles and microemulsion droplets~\cite{Milner87}. 
In their work, however, the membrane was assumed to be an incompressible 2D sheet immersed in 
fluids having the same viscosity on either side of the membrane.
This case is equivalent to a bilayer that moves together as a single entity with no difference in the number 
of molecules in the upper and lower leaflets, i.e. $\phi_0^\Delta \to 0$.
The bending relaxation rate obtained by Milner and Safran corresponds to a limiting case of 
Eq.~(\ref{eq:limit1}) when $E=1$.
The slipping relaxation rates for planar bilayers was originally discussed by Seifert and Langer who 
included the dissipation due to inter-monolayer friction~\cite{Seifert93}.
As mentioned before, all the results obtained for the planar membrane case can be reproduced from 
our results by setting $n = q r_0$ ($q$ being a wavenumber) and taking the limit of $r_0 \to \infty$.

It is worthwhile to mention that, even though we have discussed only the case when the vesicle size is 
$r_0=10$~$\mu$m, our results can be readily used to predict the relaxation behavior of vesicles 
of other sizes, provided that the parameters $k'$, $\sigma'$, $\lambda'$, and $w'$ are scaled 
appropriately. 
For smaller vesicles such as $r_0 < 10$~$\mu$m, the physically relevant $n$-range is reduced, 
and even small changes in viscosity contrast significantly affects the relaxation behavior.
Vesicles in biological systems usually belong to this size range.

The effects of surface tension on the relaxation dynamics of membranes were investigated in 
some of the previous works~\cite{Fournier15,Okamoto16}. 
In general, the surface tension makes the small-$n$ relaxation of the bending mode faster.
Arroyo \textit{et~al.}\ studied the dynamics of fluid membranes with general curved shapes, 
not restricted to a quasi-spherical vesicle, and also a membrane with free or internal 
boundaries~\cite{Arroyo09}. 
They took into account the 2D viscosity of the membrane monolayers, which has been neglected 
in our work.
Based on the previous results, we expect that including the membrane 2D viscosity will slow down 
only the large-$n$ slipping relaxation modes~\cite{Seifert93,Arroyo09}.

As a future work, it is interesting to consider the situation where the internal material is a viscoelastic 
fluid as investigated in the experiment~\cite{Viallat04}.
Although, the viscoelasticity of the membrane itself has been taken into account in some of the previous 
works~\cite{Levine02,Rochal05}, the viscoelasticity of the fluid inside has not yet been studied.
One can expect that the existence of different timescales due to the viscoelasticity of the inner fluid
will make the dynamics of the vesicle much richer~\cite{Komura12a,Komura12b,Komura15}.

\acknowledgments

T.V.S.K.\ thanks Tokyo Metropolitan University for the support provided through the 
co-tutorial program.
S.K.\ acknowledges support from the Grant-in-Aid for Scientific Research on
Innovative Areas ``\textit{Fluctuation and Structure}" (Grant No.\ 25103010) from the Ministry
of Education, Culture, Sports, Science, and Technology of Japan,
the Grant-in-Aid for Scientific Research (C) (Grant No.\ 15K05250)
from the Japan Society for the Promotion of Science (JSPS).

\appendix
\section{Rayleighian functional}
\label{appa}

The complete expression for the Rayleighian functional used to derive the basic equations 
in Sec.~\ref{sec:basic_equations} is given by,
\begin{widetext}
\begin{align}
\mathcal{R} &= \sum_{\epsilon = \pm} \int_{\epsilon} {\rm d}V \, \left[ \eta^\epsilon G^{\sigma \alpha} G^{\rho \beta} D_{\alpha \beta}^\epsilon D_{\sigma \rho}^\epsilon - P^\epsilon G^{\sigma \alpha} \partial_\alpha V_\sigma^\epsilon \right]  + \sum_{\epsilon = \pm} \int {\rm d}A_0 \, \left[ g^{ij} \mu_i^\epsilon \left( v_j^\epsilon - V_j^\epsilon \right) + \xi^\epsilon \left( r_0\frac{\partial u}{\partial t} - V_{(r)}^\epsilon \right) + \frac{w}{2} (v_i^+ - v_i^-)^2 \right] \nonumber \\
  & + \int {\rm d}A_0 \, \zeta^\Delta \left[ \frac{\partial}{\partial t}\delta \phi^\Delta + \frac{1}{2} \left(g^{ij}\partial_jv_i^+ - g^{ij}\partial_jv_i^-\right) \right] + \int {\rm d}A_0 \, \zeta^{\Sigma} \left[\frac{\partial}{\partial t}\delta \phi^{\Sigma} + \frac{1}{2} \left(g^{ij}\partial_jv_i^+ + g^{ij}\partial_jv_i^-\right) + 2 \frac{\partial u}{\partial t}\right] \nonumber \\
  & + \int {\rm d}A_0 \, \Bigg[ \Big(-k \left(\phi_0^\Delta\right)^2 \left(2u + \nabla_\perp^2 u \right) + 4k\phi_0^\Delta \delta \phi^{\Delta}  + \frac{\kappa}{r_0^2} \left( 2 \nabla_\perp^2 u + \nabla_\perp^4 u \right) + \frac{2 \lambda}{r_0} \phi_0^\Delta \nabla_\perp^2 u + \frac{4 \lambda}{r_0} \phi_0^\Delta u - \frac{2 \lambda}{r_0} \delta \phi^{\Delta} \nonumber \\
  & - \sigma (2u + \nabla_\perp^2u) + \frac{\lambda}{r_0} \nabla_\perp^2 \delta \phi^\Delta \Big) \frac{\partial u}{\partial t} + \Big( 2k \delta \phi^\Delta + 4k\phi_0^\Delta u - \frac{\lambda}{r_0}(2u - \nabla_\perp^2 u) \Big) \frac{\partial }{\partial t}\delta \phi^\Delta + \Big( 2k \delta \phi^\Sigma \Big) \frac{\partial}{\partial t}\delta \phi^{\Sigma} \Bigg].
\label{eq:rayleighian}
\end{align}
\end{widetext}

\section{Force balance equations}
\label{appb}

In this Appendix, we derive a set of four linear equations to eliminate the variables introduced in the 
solution of Stokes equations.
These are the force balance equations and the boundary conditions at the membrane surface.
We expand all the variables in terms of the spherical harmonics, and equate the components of $Y_{nm}$.
We use Eqs.~(\ref{eq:mu_elim}), (\ref{eq:xi_elim}), (\ref{eq:zeta_delta_elim}), and (\ref{eq:zeta_sigma_elim}) for eliminating the Lagrange multipliers.

Using the boundary condition Eq.~(\ref{eq:bc_impenetrable}), we can write 
\begin{align}
& \frac{-(n + 1)}{r_0} \psi_{nm}^+ + \frac{n+1}{2\eta^+(2n - 1)} r_0  P_{nm}^+  \nonumber \\
&= \frac{n}{r_0} \psi_{nm}^- + \frac{n}{2\eta^-(2n + 3)} r_0 P_{nm}^{-}  \nonumber \\
&  =  r_0\frac{\partial}{\partial t}u_{nm}. \label{eq:impenetrable_sh}
\end{align}
The force balance along the radial direction in Eq.~(\ref{eq:ldot_min}) expressed in the spherical harmonics is
\begin{align}
& -\frac{2 \eta^+(n+1)(n+2)}{r_0} \psi_{nm}^+ + \frac{2 \eta^- n (n-1)}{r_0} \psi_{nm}^-  \nonumber \\
& + \left( \frac{n(n+1)}{2n-1} + 1 \right) r_0 P_{nm}^+ + \left(\frac{n(n+1)}{2n+3} - 1 \right) r_0 P_{nm}^- \nonumber \\
  &+ \left[4k\phi_0^\Delta - \frac{2 \lambda}{r_0} - \frac{\lambda}{r_0} n(n+1)\right] \delta \phi_{nm}^\Delta - 4 k \delta \phi_{nm}^\Sigma \nonumber \\    & + \left( k \left(\phi_0^\Delta\right)^{2} + \frac{\kappa}{r_0^2} n(n+1) - \frac{2 \lambda}{r_0} \phi_0^\Delta + \sigma \right) \nonumber \\
  &\, \times \left[n(n+1)-2\right] u_{nm} = 0.
  \label{eq:radial_force_balance_sh}
\end{align}
The lateral force balance along the $\theta$-direction yields
\begin{align}
& \frac{k}{r_0} \left(\delta \phi_{nm}^\Sigma + \delta \phi_{nm}^\Delta\right) + \frac{\eta^+}{r_0^2} \left(2n + 4 + \frac{wr_0}{\eta^+} \right) \psi_{nm}^+ \nonumber \\
& + \left[ \frac{2k}{r_0} \phi_0^\Delta - \frac{2 \lambda}{r_0^2} - \frac{\lambda}{2r_0^2} \left(n(n+1) - 2\right) \right] u_{nm} \nonumber \\
& + \left[ - \frac{(n-2)}{2n(2n-1)}\left( 1 + \frac{wr_0}{\eta^+} \right) - \frac{1}{2} \right] P_{nm}^+  - \frac{w}{r_0} \psi_{nm}^- \nonumber \\
& - \frac{wr_0}{\eta^-} \frac{(n+3)}{2(n+1)(2n+3)} P_{nm}^- = 0,
\label{eq:lateral_force_balance_plus_sh}
\end{align} 
and
\begin{align}
& \frac{k}{r_0} \left(\delta \phi_{nm}^\Sigma - \delta \phi_{nm}^\Delta\right) + \frac{\eta^-}{r_0^2} \left(2n - 2 + \frac{wr_0}{\eta^-} \right) \psi_{nm}^-  \nonumber \\
& - \left[ \frac{2k}{r_0} \phi_0^\Delta - \frac{2 \lambda}{r_0^2} - \frac{\lambda}{2r_0^2} \left(n(n+1) - 2\right) \right] u_{nm} \nonumber \\
&  + \left[ - \frac{(n+3)}{2(n+1)(2n+3)}\left( 1 - \frac{wr_0}{\eta^-} \right) - \frac{1}{2} \right] P_{nm}^-  \nonumber \\
&- \frac{w}{r_0} \psi_{nm}^+ + \frac{wr_0}{\eta^+} \frac{(n-2)}{2n(2n-1)} P_{nm}^+ = 0,
\label{eq:lateral_force_balance_minus_sh}
\end{align}
for the outer and inner monolayers, respectively.

Equations~(\ref{eq:impenetrable_sh}), (\ref{eq:radial_force_balance_sh}), 
(\ref{eq:lateral_force_balance_plus_sh}), and (\ref{eq:lateral_force_balance_minus_sh}) constitute a 
set of four simultaneous linear equations, which can be used to eliminate $\psi_{nm}^{\pm}$ and $P_{nm}^{\pm}$. 
Then we obtain the dynamical equations for the variables $u$, $\phi^\Delta$, and $\phi^\Sigma$.

\section{Elements of the matrix $\mathbf{C}$}
\label{appc}

Here, we list the elements of the symmetric matrix $\mathbf{C}$ defined in 
Eq.~(\ref{eq:full_dynamical_matrix}).
\begin{align}
  &C_{11} = \frac{2n(n+1)}{D} (2n+1)\left[w'(1+E)+(2n+1)E\right],
\end{align}
\begin{align}
  &C_{12} = C_{21} = - \frac{3n(n+1)}{D} (2n+1)E, 
\end{align}
\begin{align}  
  C_{13} = C_{31} &=  - \frac{2n(n+1)}{D} \nonumber \\
 & \times \Big[ w'\left[(n+1)+(n-1)E\right] - (2n+1)^2E \Big], 
 \end{align}
 \begin{align}
  C_{22} &= \frac{n(n+1)}{D}\Bigg[ (n^2-1)(2n^2+4n+3)E^2 \nonumber \\
  &+ E\left(\frac{3}{2} + 2n\left(2n^3 + 4n^2+n-1\right)\right) \nonumber \\
  & + n(n+2)(2n^2+1) \Bigg], 
  \end{align}
  \begin{align}
  C_{23} = C_{32} &= \frac{n(n+1)}{D} \Bigg[(n^2-1)(2n^2+4n+3)E^2 \nonumber \\
  &+ \frac{9}{2}(2n+1)E - n(n+2)(2n^2+1)\Bigg], 
  \end{align}
  \begin{align}
  C_{33} &= \frac{n(n+1)}{D} \Big[(n^2-1)(2n^2+4n+3)E^2\nonumber \\
  & + 2w'(n-1)(2n^2+5n+5)E + \left[5-8n^2(n+1)^2\right]E  \nonumber \\
  &
 +(n+2)\left[2w'(n(2n-1) + 2)+n(2n^2+1)\right]\Big].
  \end{align}
The common denominator in all the components is
\begin{align}
D &= 2 \Bigg[ 2E^2(n-1)(n+1)(2n^{2}+4n+3)\left[ w' + (2n+1) \right] \nonumber \\
& + E \Big[ w'(8n^4+16n^3+4n^2-4n+3) \nonumber \\
&+ 2n(n+2)(2n+1)(2n^{2}+1) \Big] \nonumber \\
&+ 2w' n(n+2)(2n^{2}+1)\Bigg].
\end{align}


\end{document}